\documentclass[useAMS, usenatbib, usegraphicx]{mn2e}
\usepackage{amssymb}
\usepackage{booktabs}
\newcommand\ApJ{ApJ}

\newcommand\ApJS{ApJS}
\newcommand\AnA{A\&A}

\def\alf{Alfv\'en\,}
\def\alfc{Alfv\'enic\,}
\def\bq{\begin{equation}}
\def\eq{\end{equation}}
\def\ee #1 {\times 10^{#1}}
\def\ut #1 #2 { \, \rmn{#1}^{#2}}
\def\u #1 { \, \rmn{#1}}

\let\grad=\nabla
\newcommand\cross{\bmath{\times}}
\newcommand\etaH{\eta_\mathrm{H}}

\def\curl{{\grad \cross}}
\def\div #1 {\grad \cdot #1}
\def\veps{\varepsilon}

\def\hB{\hat{\bmath b}}
\def\rhon{\rho_{in}}
\def\rhox{\rho_{ex}}

\def\v{\bmath{v}}
\def\dv{\delta \v}

\def\U{\bmath{U}}
\def\v{\bmath{v}}
\def\vi{\bmath{v}_i}

\def\ve{\bmath{v}_e}
\def\vi{\bmath{v}_i}
\def\vB{\bmath{v}_B}
\def\vn{\bmath{v}_n}
\def\va{v_A}
\def\J{\bmath{J}}
\def\Jpa{\bmath{J_\parallel}}
\def\B{\bmath{B}}
\def\dB{\delta \B}

\def\E{\bmath{E}}
\def\dE{\delta \E}
\def\dEx{\delta E_x}
\def\dEy{\delta E_y}

\def\J{\bmath{J}}

\def\drho{\delta\rho}
\def\dv{\bmath{\delta\v}}
\def\dvx{\delta v_x}
\def\dvy{\delta v_y}
\def\dE{\bmath{\delta\E}}
\def\dB{\bmath{\delta\B}}

\newcommand{\delt} [1] {\frac{\partial #1}{\partial t}}

\def\oma{{\omega_A}}
\def\b1{{\bar{\omega}}}
\def\bo2{\bar{\omega}^2}
\def\L1{{\bf {\it L}}}

\title{Surface Wave Propagation in non--ideal plasmas}
\author[B. P. Pandey and C. B. Dwivedi]{B. P. Pandey$^{1}$
\thanks{E-mail:birendra.pandey@mq.edu.au;jagatpurdwivedi@gmail.com}, and C. B.  Dwivedi$^2$
\\
$^{1}$Department of Physics, Macquarie University, Sydney 2109, Australia\\
$^2$Ved--Vijnanam Pravartanam Samitihi, Pratapgarh (Awadh),  Jagatpur 230306, Bharat (India)}
\date{\today}
\pagerange{\pageref{firstpage}--\pageref{lastpage}}
\pubyear{2014}
\begin{document}
\maketitle
\label{firstpage} 
\begin{abstract}
\maketitle 
The properties of surface waves in a partially ionized, compressible magnetized plasma slab are investigated in this work. The waves are affected by the non—-ideal magnetohydrodynamic (MHD) effects which causes finite drift of the magnetic field in the medium. When the magnetic field drift is ignored, the characteristics of the wave propagation in a partially ionized plasma fluid is similar to the fully ionized ideal MHD except now the propagation properties depend on the fractional ionization as well as on the compressibility of the medium. 

The phase velocity of the sausage and kink waves increases marginally (by a few percent) due to the compressibility of the medium in both {\it ideal} as well as Hall—diffusion--dominated regimes. However, unlike {\it ideal} regime, only waves below certain cut—-off frequency can propagate in the medium in Hall dominated regime. This cut-—off for a thin slab has a weak dependence on the plasma beta whereas for thick slab no such dependence exists. More importantly, since the cut--off is introduced by the Hall diffusion, the fractional ionization of the medium is more important than the plasma compressibility in determining such a cut—off.  Therefore, for both compressible as well incompressible medium, the surface modes of shorter wavelength are permitted with increasing ionization in the medium.  We discuss the relevance of these results in the context of solar photosphere--chromosphere.
\end{abstract}


\section{Introduction}
The matter in the Universe is in a partially ionized plasma state with varying degree of ionization determined locally by ambient thermodynamic conditions.  The star--forming molecular clouds, accretion discs, solar atmosphere, planetary rings, cometary tail, Earth$\textquoteright$s ionosphere, and laboratory plasma devices are some of the examples of a partially ionized medium. The presence of a magnetic field and collision in addition, introduces bewildering variety of scales over which dynamics of such a gas manifests itself: as radiation from infalling matter in black holes; as protostellar and protoplanetary nebula ensconced in dusty discs; as million degree solar corona; as majestic northern lights and finally, our own effort to tap the endless source of fusion energy.  The collision and the magnetic field often have opposite effect on the plasma dynamics: whereas the magnetic field {\it freezes / confines} the charged particle motion, collision often causes diffusion. The neutral gas in such a partially ionized mixture can be acted upon by the magnetic torque via collisional momentum exchange with the plasma particles. All in all, interplay between the collision and magnetic field may cause relative drift between various constituents of a partially ionized gas. To quantify, the relative drift between the plasma and neutrals causes ambipolar diffusion while relative drift between the electrons and ions causes Hall diffusion. The drift between the electrons and neutrals causes Ohm diffusion. Clearly, dynamics of a partially ionized gas, in general, cannot be described in the framework of ideal magnetohydrodynamics (MHD) and non--ideal effects such as Ohm, ambipolar and Hall becomes important. 

In a weakly ionized medium, neglecting plasma inertia, a linear relationship between the electric field $\E$ and plasma current $\J$ can be easily derived $\E = \bmath{\eta}\cdot \J\,,$ where $\bmath{\eta}$ is the diffusivity tensor. However, this approach is not very fruitful in situations where plasma inertia and non–-ideal MHD effects are simultaneously important i.e. when the plasma is not {\it weakly} but {\it partially} ionized. Typical example includes a solar photosphere–-chromosphere transition region, Earth$\textquoteright$s ionosphere,  protoplanetary discs, discs around cataclysmic variables, etc. The collisional, non--ideal MHD effects can be included in the dynamics either in the multi-–fluid or the single--fluid framework. Whereas multi–-fluid framework is well suited to describe the high frequency fluctuations in the medium, the single--fluid framework is often used to describe the low frequency dynamics of the medium. Given the complexity of multi—-fluid dynamics, it is more than a passing curiosity to reduce the multi--fluid description to simplified, single—-fluid, MHD like description. However, the validity of single--fluid MHD like description is tied to the fractional ionization and ion-—neutral collision frequency since the dynamical frequency of interest $\omega$ must satisfy \citep{P06, P08a}
\bq
\left(\frac{\omega}{\nu_{in}}\right) \lesssim X_e^{-1/2} \equiv \left(\frac{n_e}{n_n}\right)^{-1/2}\,,
\label{eq:cond}
\eq  
where $\nu_{in}$ is the ion-—neutral collision frequency and $n_e\,,n_n$ are the electron and neutral number densities. The multi—-fluid description \citep{V08, Z12, S13, K14} is the only way out when the frequencies are higher than that given by Eq.~(\ref{eq:cond}).

\begin{figure}
\includegraphics[scale=0.30]{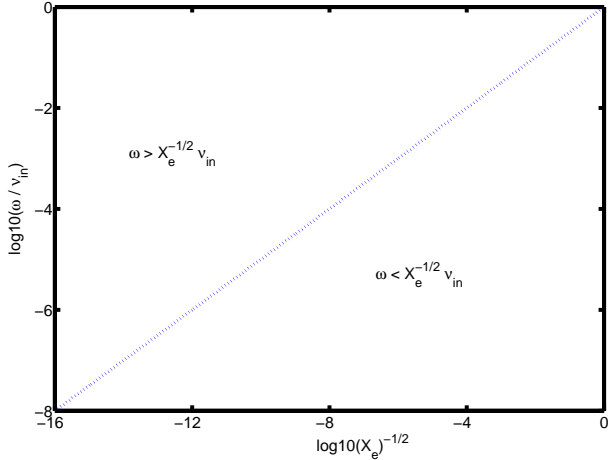}
\caption{\large{We plot $\omega / \nu_{in}$ against fractional ionization in the above figure.}}
\label{fig:fig1}
\end{figure}
In Fig.~(\ref{fig:fig1}) $\omega / \nu_{in}$ is plotted against fractional ionization. In Earth$\textquoteright$s E-region, where $X_e \lesssim 10^{-12}$, ultra low frequency events are amenable to single--fluid description whereas in the solar photosphere—-chromosphere where $X_e \lesssim 10^{-4}$, relatively higher frequency fluctuations can be described in the single--fluid framework. Therefore, the quantification {\it high} or, {\it low} frequency depends on the local physical conditions. 

The transition from a weakly to fully ionized region poses considerable challenge to the modelling of a partially ionized gas. Although such a framework was formally developed more than half a century ago \citep{C57, B65}, the validity of multi fluid framework, Eq.~(\ref{eq:cond}) and associated spatial and temporal non-—ideal MHD scales were clearly elucidated only recently \citep{P06, P08a}. For example Hall diffusion becomes dynamically important at a much lower frequency than the limit derived from the fully ionized Hall MHD. Defining $\rho_{i\,, n} = m_{i\,,n}\,n_{i\,,n}$ as the ion (neutral) mass density where $m_{i\,,n}$ is the ion (neutral) mass and $n_{i\,,n}$ is the ion (neutral) number density, $\rho = \rho_i + \rho_n$ as the bulk mass density and $\omega_{ci} = e\,B / m_i\,c$ as the ion--cyclotron frequency where $e\,,B$ and $c$ are the electron charge, magnetic field and speed of light, respectively, the Hall frequency $\omega_H $ \citep{P08a} 
\bq
\omega_H = \frac{\rho_i}{\rho}\,\omega_{ci}\,,
\label{eq:hfy}
\eq
suggests that in a weakly ionized medium ($\rho_i \rightarrow 0$), Hall frequency becomes negligible. Therefore, unlike ideal MHD, where Hall effect is important only when the dynamical frequency $\omega$ is of the order of or, larger than the ion--cyclotron frequency, i.e., $\omega \gtrsim \omega_{ci}$, in a weakly ionized medium Hall effect becomes dynamically important for all frequencies of the order of or, above Hall, i.e. $\omega \gtrsim \omega_H$ which is easily satisfied since $\omega_H \approx 0$.  Clearly, very low frequency fluctuations will be affected by the Hall diffusion of the magnetic field in a weakly ionized medium. Therefore, it is not surprising that Hall plays important role in the angular momentum transport in  protoplanetary discs \citep{W99}, in destabilizing the solar flux tubes \citep{PW12, PW13}, in the Earth$\textquoteright$s ionosphere, etc. 

Defining $v_A = B / \sqrt{4\,\pi\,\rho}$ as the \alf velocity in the bulk medium, we see that the Hall scale $L_H$ \citep{P08a} 
\bq
L_H = \left(\frac{\rho_i}{\rho}\right)^{1/2}\,\frac{v_A^2}{\omega_H}\,,
\eq 
becomes very large in the $\omega_H \rightarrow 0$ limit, i.e. in a weakly ionized plasma Hall operates over a large scale. This feature of a weakly ionized medium  becomes all the more remarkable if we recall that in a fully ionized medium Hall scale is generally very small ($\sim$ ion skin depth) and thus, the Hall effect is inconsequential over large scales. 

The MHD waves are capable of carrying energy millions of miles away from their source of origin and thus, play an important role in the space and laboratory plasmas. For example, these waves may provide efficient heating to both fusion \citep{T73, C74, K77, M92, P95} as well as solar coronal plasmas \citep{I78, G94, P97a, P97b, P98, A09, G11}. The \alf waves are possible source of turbulence in molecular clouds \citep{PV07}. The past study on wave propagation and resonance heating is largely confined to the fully ionized medium. Only recently the investigation of wave propagation in a partially ionized medium has picked up momentum due to its importance to space weather \citep{G98, K02, KG06, K06, P08b, S09, S10, G11, K12, Z12, P13, S13, K14}.  The drift of magnetic field through the matter in a partially ionized gas provides new pathways through which energy can be channelled to waves \citep{PW12, PW13}. 

The most important development in the study of MHD waves over the last decade in space plasmas has been numerous observational claims of MHD waves in the solar atmosphere [Goossens et al. (2013) and references therein]. This has triggered important theoretical development towards explaining various observations. However, basic MHD model, which is employed to explain these observations assume the presence of fully ionized matter, a far cry from the state of matter below the chromosphere—-corona transition region. However, since the presence of a magnetic field and density inhomogeneity considerably complicates the investigation of wave propagation, the neglect of neutral—-plasma interaction, although not ideal, is a good starting point. Therefore on the theoretical front challenge is how to generalize the existing MHD wave propagation model to incorporate non—-ideal, neutral—-plasma collision dominated effects. This development is crucial for our understanding of underlying physical processes at the transition region and thus for the resonant heating of the corona and solar wind launching.

The Sun holds a special place in the investigation of MHD waves. The observational validation of various wave modes in our backyard allows us to spread our wings to distant stars and interstellar medium. Thus it is not at all surprising that the large effort has been dedicated to understand various wavelike features in the solar atmosphere in the framework of  both ideal MHD and Hall MHD \citep{P72, R79, W79, E82, C85, C86, G94, Z96, G09, Z09}. This work will primarily focus on understanding the nature of wave propagation in a partially ionized compressible medium and choose as a concrete example solar atmosphere as possible application of the results. Recent investigation of the surface wave in cylindrical filaments \citep{S09, S10} suggests the important role of non—-ideal MHD effects for short wavelength fluctuations. We build upon the past studies of the surface waves in ideal and Hall MHD and generalize it to a partially ionized, compressible medium.  

The paper is organized in the following fashion. The basic framework to investigate the surface waves is given in Sec. 2. The general dispersion relation for sausage and kink mode is derived and various limiting cases are discussed in section 3. In Sec.~4 discussion and a brief summary of the results is presented and future direction is indicated.
\section{Basic model}
We shall assume a partially ionized plasma consisting of electrons, singly charged ions and neutral particles. 
The single--fluid MHD—-like description of such a gas is given by following set of equations \citep{P08a}
\bq
\frac{\partial \rho}{\partial t} + \grad\cdot\left(\rho\,\v\right) = 0\,,
\label{eq:cont}
\eq
\bq
\rho\,\frac{d\v}{dt}=  - \nabla\,P + \frac{\J\cross\B}{c}\,,
\label{eq:meq}
\eq
where $ \v = (\rho_i\,\vi + \rho_n\,\vn)/\rho$, $\J = n_e\,\left(\vi - \ve\right)$ is the current density, $\B$ is the magnetic field and $P = P_e + P_i + P_n$ is the total pressure.
The induction equation is 
\bq
\delt \B = \curl\left[
\left(\v + \v_B\right)\cross\B - \frac {4\,\pi\,\eta}{c}\,\Jpa\right]\,,
\label{eq:indA}
\eq
where the magnetic drift velocity ($\vB$) is defined as 
\bq
\vB = \eta_P\,\frac{\left(\grad\cross\B\right) \cross\hB}{B} -– 
\eta_H\,\frac{\left(\grad\cross\B\right)_{\perp} }{B}\,, 
\label{eq:md0}
\eq
with $\hB = \B /B$, $\Jpa = (\J\cdot \hB)\,\hB$, $\left(\grad\cross\B\right)_{\perp} = \grad\cross\B -- \left(\hB\cdot \grad\cross\B\right)\,\hB$ and $\eta_P = \eta_A + \eta$ is Pedersen diffusion. The Ohm ($\eta$), ambipolar ($\eta_A$) and Hall ($\etaH$) diffusivities are given as
\bq \eta =
\frac{c^2}{4\,\pi\sigma}\,\,, 
\eta_A = \left(\frac{\rho_n}{\rho_i}\right)\,\frac{v_A^2}{\nu_{n\,i}}\,, \mbox{and}\, \etaH = \frac{v_A^2}{\omega_H}\,.
\label{eq:diffu}
 \eq
Here $\sigma = c\,e\,n_i \,\left( \beta_e + \beta_i \right) / B$ is the parallel conductivity defined in terms of plasma Hall parameter 
\bq
\beta_j = \frac{\omega_{c\,j}} {\nu_{j\,n}}\,, 
\eq
which is a ratio between the plasma—-cyclotron ($\omega_{cj} = e_j\,B / m_i\,c$ with $e_j = \pm e$) and plasma-—neutral ($\nu_{jn} = \rho_n\,\nu_{nj} / \rho_j$) collision frequencies.

To close the above--mentioned set of equations an adiabatic equation of state $P / \rho^{\gamma} = \mbox{const}.$ will be assumed. Here $\gamma$ is the ratio of two specific heats. Since the background is spatially uniform, the spatial derivative of $\rho_0$ and $P_0$ vanish. Here subscript $0$ is used to denote background quantities. Thus writing pressure and density as $f = f_0 + \delta f$ where the fluctuation $\delta f$ is much smaller than $f_0$ we see that in the adiabatic case
\bq
\frac{P}{\rho^{\gamma}} \approx \frac{P_0}{\rho_0^{\gamma}}\,\left(1 + \frac{\delta P}{P_0} - \gamma\,\frac{\delta \rho}{\rho_0}\right) = \frac{P_0}{\rho_0^{\gamma}}\,,
\eq
and thus we may write $\delta P = c_s^2\,\delta \rho$ where the sound speed $c_s = \sqrt{\gamma\,P_0 / \rho_0}$. We shall use this equation of state below to eliminate perturbed pressure in terms of perturbed density. Further, equilibrium quantities will be denoted without subscript zero below.   

We shall consider a partially ionized, slab of thickness $2\,x_0$ 
having following piecewise constant density and pressure
\bq
\rho(x) = \left\{
\begin{array}{ll}
\rhon\,,P_{in} & \mbox{if $|x|  \leq x_0$};\\
\rhox \,,P_{ex}& \mbox{if $|x| > x_0$}\,, 
\end{array}
\right.
\eq 
threaded by uniform magnetic field $\B = B \,\hat{\bmath{z}}$ parallel to the undisturbed surface of tangential discontinuity. Often a simplified model of the solar flux tube is modelled by such a plasma slab with piecewise constant density and pressure \citep{E82}. However, such a discontinuity fundamentally changes the behaviour of the \alfc vorticity propagation in the medium. Whereas in an infinite uniform medium, the \alfc vorticity in nonzero in the entire volume, the density jump confines the vorticity to the surface layer $x = x_0$ only \citep{G12}.  

We shall further assume that magnetic flux is frozen in the plasma slab, moving with $\v + \vB$, i.e. neglect last term in the induction equation, (\ref{eq:indA}). This implies that either the field aligned parallel current is zero, i.e. $\Jpa = 0$ or, the Ohm diffusion is unimportant.
 The linearized equations are
\bq
\delt\drho + \rho\,\nabla\cdot\dv = 0\,.
\label{eq:Lc}
\eq
\bq
\delt\dv + \nabla\left(c_s^2\frac{\drho}{\rho} + \frac{\B\cdot\dB}{4\,\pi\,\rho}\right) = \frac{\left(\B\cdot\nabla\right)\dB}{4\,\pi\,\rho} 
\label{eq:Lm}
\eq
\bq
\delt\dB = \left(\B\cdot\nabla\right)\left(\dv  + \delta \vB\right) - \B\,\nabla\cdot \left(\dv  + \delta \vB\right)\,.
\label{eq:Li}
\eq
Fourier analysing the perturbed quantities as $exp{\left(-i\,\omega\,t + i\,k\,z\right)}$, and defining 
${\bf{L}} = \frac{d^2}{dx^2} -– k^2\,,\, \mbox{and}\, \omega_A^2 = k^2\,v_A^2$, the momentum and induction equations can be written in the following form respectively. 
\begin{eqnarray}
\left(
\begin{array}{cc} \omega + \frac{\omega\,c_s^2}{\left(\omega^2-—k^2\,c_s^2\right)}\frac{d^2}{dx^2}   &  0\\
                   0       & \omega
  \end{array}
\right)\,\left(\begin{array}{c} \delta v_x\\
                                \delta v_y    
\end{array}
\right)
\nonumber\\ 
= v_A^2
\left(
\begin{array}{cc} \frac{1}{k}{\bf{L}}   &  0\\
                   0       & - k
  \end{array}
\right)\, \left(\begin{array}{c} \delta B_x / B\\
                                \delta B_y / B    
\end{array}
\right)\,,
\label{eq:lin1}
\end{eqnarray} 
and,
\begin{eqnarray}
\left(
\begin{array}{cc} \omega –- i\,\eta_P\,{\bf{L}}   &  0\\
                   i\,\eta_{H}\,{\bf{L}} & \omega + i\,k^2\,\eta_{P}
  \end{array}
\right)\,\left(\begin{array}{c} \delta B_x / B\\
                                \delta B_y / B    
\end{array}
\right)
\nonumber\\
 = k\left(
\begin{array}{cc} - 1   &  i\,\frac{\eta_H}{v_A^2}\,\omega\\\
                   0       & - 1
  \end{array}
\right)\, \left(\begin{array}{c} \delta v_x\\
                                \delta v_y    
\end{array}
\right)\,.
\label{eq:lin2}
\end{eqnarray}
While writing the above--mentioned equations, we have assumed diffusivities $\eta_A$ and $\eta_H$ to be constant. Since investigation of surface wave in a plasma slab is the concern of this work, such an assumption will not impact the results. 

Defining $a = \oma^2 / \omega^2\,,$ $\beta = c_s^2 / \va^2$ and 
\begin{eqnarray}
A =  1 - \frac{1}{a\,\beta}\,,q^2 = A\,k^2\,\Big[1 - \frac{1}{1 + \beta\,\left(1 - a \right)}\Big]\,, 
\nonumber\\
F = \left(\frac{i\,A\,\omega}{\eta_H}\right) \left[ 1 -– a + i\,\frac{k^2\,\eta_P}{\omega}\right]\,,
\nonumber\\
F_1 = \frac{\eta_P}{\eta_H} \frac{q^2}{k^2\,\left(1 –- a\right)}
\left[1 - a + i\,\frac{k^2\,\eta_P}{\omega}\left(1 + \frac{\eta_H^2}{\eta_P^2}\right) \right]\,, 
\label{eq:FDF}
\end{eqnarray}
Eqs.~(\ref{eq:lin1}) and (\ref{eq:lin2}) can be written in a compact form as
\begin{eqnarray}
\left(\frac{d^2}{dx^2} –- q^2\right)\dvx
+ F_1\left(\frac{d^2}{dx^2} –- k^2\right)\dvy = 0\,,
\label{eq:xx1}
\\
\left(\frac{d^2}{dx^2} –- k^2\,A\right)\dvx –- F\,\dvy = 0\,.
\label{eq:xx2}
\end{eqnarray}
Note that in the incompressible limit, i.e. when $\beta \rightarrow \infty$  Eqs.~(\ref{eq:xx1}) and (\ref{eq:xx2}) reduces to Eqs.~(20) and (21)  of \cite{P13} except for a typo $i$ in their $F_1$ in equation (22). 

In the absence of non—-ideal MHD effect, i.e. setting $\eta_P = \eta_H = 0$, Eqs.~(\ref{eq:xx1}) and (\ref{eq:xx2}) reduces to the following equation 
\bq
\left(\frac{d^2}{dx^2} –- q^2\right)\dvx = 0\,,
\eq 
which is Eq.~(5) of \cite{E82} except now we are dealing with a partially ionized medium. Clearly, low frequency surface wave can propagate undamped in the partially ionized medium \citep{U98}. 

In the absence of Hall diffusion, Eqs.~(\ref{eq:xx1}) and (\ref{eq:xx2}) reduces to the following uncoupled equations for $\dvx$ and $\dvy$ components
\begin{eqnarray}
\Bigg[\frac{i\,a\,\eta_P}{\beta\,\omega}\,{\bf{L}}^2 - \left(1-\frac{i\,a\,k^2\,\eta_P}{\beta\,\omega}\right)\, {\bf{L}} -– k^2\Bigg]\dvx = 0\,
\nonumber\\
\left( \omega^2 - \omega_A^2 + i\,k^2\,\eta_P\,\omega\right)\dvy = 0\,. 
\label{eq:pamb}
\end{eqnarray}
In the incompressible ($\beta \rightarrow \infty$) limit, the above--mentioned equations are identical to  Eq.~(55) of \cite{P13}. In $\beta \rightarrow 0$ limit, from $\dvx$ equation above we may write $\dvx = Q\,x + Const.$. Since fluctuation decays over Pedersen diffusion time--scale ($\sim 1 / k^2\,\eta_P$), the physical solution requires that $const. = - Q\,x_0$ at the surface boundary. Thus, we may conclude like an incompressible case that in a purely Pedersen regime the waves will disappear at the surface boundary over $\sim 1 / k^2\,\eta_P$.  There is no effect of compressibility on $\dvy$ equation which also describes wave damping at a rate $k^2\,\eta_P$.  To summarize, notwithstanding the compressibility of the fluid, surface waves are always damped by Pedersen diffusion. 

In a homogeneous plasma, we may Fourier analyse the $x-$dependence as $\exp(inx)$ and the above--mentioned equations reduce to the following dispersion relation
\begin{eqnarray}
\left(\frac{1}{\beta} + 1 -– a \right)\,\left(1 -–  a + i\,\frac{k^2\,\eta_P}{\omega}\right) + i\,\frac{k^2\,\eta_P}{\omega}\,
\left(\frac{k^2 + n^2}{n^2 + q^2}\right)
\nonumber\\
\times
\left(\frac{n^2}{k^2} + 1 - \frac{1}{a\,\beta}\right)
\left[1 –-  a + i\,\frac{k^2\,\eta_P}{\omega}\left( 1 + \frac{\eta_H^2}{\eta_P^2}\right)\right] = 0\,.
\label{eq:drn}
\end{eqnarray}
The dispersion relation can be solved numerically. However, we shall analyse it analytically in various limiting cases. 

In the incompressible limit when $\beta \rightarrow \infty$, we get from Eq.~(\ref{eq:FDF}) $A = 1\,,$ $q^2 = k^2$ and the above--mentioned dispersion relation reduces to Eq.~(23) of \cite{P13}. Since the dispersion relation in this limit has been already analysed previously, we shall explore $\beta \rightarrow 0$ limit i.e. when $c_s^2 \ll \va^2$. In this limit Eq.~(\ref{eq:drn}) becomes 
\begin{eqnarray}
a\,\left(1 -–  a + i\,\frac{k^2\,\eta_P}{\omega}\right) -i\,\frac{k^2\,\eta_P}{\omega}\,
\nonumber\\
\times \left(\frac{k^2 + n^2}{n^2 + q^2}\right)
\left[1 –-  a + i\,\frac{k^2\,\eta_P}{\omega}\left( 1 + \frac{\eta_H^2}{\eta_P^2}\right)\right] = 0\,,
\label{eq:drn1}
\end{eqnarray}
which in the absence of Hall ($\eta_H = 0$) gives
\bq
\left(\omega^2 + i\,k^2\,\eta_P\,\omega -–  k^2\,\va^2\right) \left(\omega^2 + i\,\chi^2\,\eta_P\,\omega -–  \chi^2\,\va^2\right) = 0\,,
\label{eq:adr}
\eq
where $\chi^2 = n^2 + k^2$. The above--mentioned dispersion relation, after setting the first bracket to zero, gives  
mode
\bq
\omega \simeq \pm k\,\va –- i\,\frac{k^2\,\eta_p}{2}\,,
\eq 
which describes the damped \alf wave.
 Setting second bracket to zero also gives similar formula except now $k$ is replaced by $\chi$  and the damping rate of the wave is $\sim \chi^2\,\eta_P$ and not $\sim k^2\,\eta_P$. 

In the absence of Pedersen diffusion ($\eta_P = 0$), Eq.~(\ref{eq:drn}) becomes 
\bq
\left(\frac{\omega}{\oma}\right)^4 - \Bigg[ 1 + \frac{\chi^2}{k^2} + \frac{\chi^2\,\eta_H^2}{v_A^2}\Bigg] \left(\frac{\omega}{\oma}\right)^2 + \frac{\chi^2}{k^2} = 0\,, 
\label{eq:HLD}
\eq
which for the short wavelength ($\omega_H \ll \omega_A$) fluctuations, describes the whistler waves in the high frequency ($\omega_A \ll \omega$) limit  
\bq
\omega \cong  \left(1 + \frac{n^2}{k^2}\right)^{1/2}\,k^2\,\eta_H\,.
\eq
The above--mentioned expression has been derived by balancing the first and second term in Eq.~(\ref{eq:HLD}) after noting that the last term in square bracket is the dominant term. Similarly, electrostatic ($\curl \dE \approx 0$) ion–-cyclotron wave in the low frequency $\omega \ll \omega_A$ limit,   
\bq
\omega \cong \omega_H\,.
\eq
is derived  by neglecting the first term in Eq.~(\ref{eq:HLD})and retaining only the dominant term in the square bracket.

In the long wavelength limit, i.e. $\omega_A \ll \omega_H$, we recover usual \alf $\omega^2 = \omega_A^2$ and magnetosonic $\omega^2 = \chi^2\,\va^2$ waves. Clearly, wave propagation in limiting cases ($\beta \rightarrow \infty$ and $\beta \rightarrow 0$) display similar behaviour and thus we should anticipate that surface waves will also display similar behaviour.

Although the method of deriving the dispersion relation for the surface wave is identical to \cite{P13}, for clarity and completeness we shall repeat those steps here as well. We start by seeking the solution of Eqns.~ (\ref{eq:xx1}) and (\ref{eq:xx2}) as
\begin{eqnarray}
\dvx &=& f \left[\exp{\left(-\alpha\,x\right)} \mp \exp{\left(\alpha\, x\right)} \right]\nonumber\\ 
\dvy &=& i\,h\, \left[\exp{\left(-\alpha\,x\right)} \mp \exp{\left(\alpha\, x\right)}\right]\,, 
\end{eqnarray} 
which leads to the following equation for $\alpha$ 
\bq
\left(\alpha^2 -– q^2\right) + \left(\frac{F_1}{F}\right) \left(\alpha^2 -– k^2\right) \left(\alpha^2 -– k^2\,A\right) = 0 \,,
\label{eq:rta}
\eq
from where in the incompressible limit we get $\alpha^2 = k^2$ and $\alpha^2 = k^2\,\left(1 - F / k^2 F_1\right)$ which are same as Eq.~(31) of \cite{P13}.  In order to keep our analytical development tractable, we shall approximate the roots of $\alpha$ as
 \bq
\alpha^2 \approx k^2\,A\,,\quad \alpha^2 \approx k^2\,\Big[1 - \frac{F}{k^2\,F_1}\Big] \,,
\label{eq:rta1}
\eq
which implies that $q^2 \simeq A\,k^2$ in Eq.~(\ref{eq:rta}) is valid only if $\beta\,\left(1-a\right) \gg 1$. Thus the validity of our analysis requires that $\omega / k > v_A\,\beta / \left(\beta - 1\right)$ which sets the limit on the compressibility of the medium.

The four roots in Eq.~(\ref{eq:rta1}) corresponds to the pair of attenuation coefficient $\left(\alpha_{in1}\,,\alpha_{in2}\right)$ inside and $\left(\alpha_{ex1}\,,\alpha_{ex2}\right)$ outside the slab. Motivated by the fact that planer or cylindrical waveguides can support kink and sausage modes, we choose the general solution of $\dvx$ and $\dvy$ as a superposition of such waves. More precisely, for the sausage wave, inside the slab ($|x| < x_0$)
\begin{eqnarray}
\dvx (x) = f_1\,\frac{\sinh(\alpha_{in1}\,x)}{ \sinh(\alpha_{in1}\,x_0)} + f_2\,\frac{\sinh(\alpha_{in2}\,x)}{ \sinh(\alpha_{in2}\,x_0)}\,\\
\dvy (x) = i\,f_1\,G_{in1}\,\frac{\sinh(\alpha_{in1}\,x)}
{\sinh(\alpha_{in1}\,x_0)} + i\,f_2\,G_{in2}\,\frac{\sinh(\alpha_{in2}\,x)}{ \sinh(\alpha_{in2}\,x_0)}
\label{eq:sw1}
\end{eqnarray} 
where 
\bq
G_{j1\,,2} = - \eta_H\,\frac{\left(\alpha_{j1\,,2}^2 -– A_{j}\,k^2\right)}{A_{j}\,\left[ \left(1 –- a_{j}\right)\,\omega + i\,k^2\,\eta_A\right]}\,.
\eq

For kink surface-–wave, similar expression for the perturbed velocities can be given by replacing $\sinh$ by $\cosh$. The solution outside plasma layer is
\begin{eqnarray}
\dvx = 
s_1\,e^{-\alpha_{ex1}\, \left(x -– x_0\right)} + s_2\,e^{-\alpha_{ex2}\, \left(x -– x_0\right)} \,, x > x_0\nonumber\\ 
\nonumber\\ 
\dvx = \beta_1\,e^{\alpha_{ex1}\, \left(x + x_0\right)} + \beta_2\,e^{\alpha_{ex2}\, \left(x + x_0\right)}\,, x < - x_0\,, 
\end{eqnarray} 
and 
\begin{eqnarray}
- i\,\dvy = s_1\,G_{ex1}\,e^{-\alpha_{ex1}\,\left(x -– x_0\right)} + s_2\, G_{ex2}\,e^{-\alpha_{ex2}\,\left(x -– x_0\right)} \,,\nonumber\\
 x > x_0\nonumber\\ 
\nonumber\\ 
-i\,\dvy = \beta_1\,G_{ex1}\,e^{\alpha_{ex1}\,\left(x + x_0\right)} + \beta_2\, G_{ex2}\,e^{\alpha_{ex2}\, \left(x + x_0\right)} \,,\nonumber\\ x < - x_0\,. 
\end{eqnarray} 
The knowledge of $\dvx$ and $\dvy$ allows us to calculate the perturbed total pressure
\begin{eqnarray}
\frac{\delta p_T}{\rho} = i\,\frac{a\,\omega}{k^2} \Bigg[
\frac{1 + \beta\,\left(1 - a \right)}{1 - \beta\,a}\,\frac{d\dvx}{dx} + 
\Bigg.
\nonumber\\
\Bigg.
\left\{\frac{\eta_P}{a\,\eta_H}\left(1 - a + i\,\frac{i\,k^2\,\eta_P}{\omega}\right) + i\,\left(\frac{k^2\,\eta_H}{\omega}\right) \right\}\frac{d\dvy}{dx}
\Bigg]\,.
\end{eqnarray}
The electric field components $\dEx$ and $\dEy$ which is required for the boundary conditions, can be derived from the generalized Ohm\textsc{\char13}s law
\bq
c\,\dE = - \left(\dv + \delta \vB\right)\cross\B\,, 
\eq
which yields
\bq
\frac{c\,\dEx}{B} = - \frac{1}{a} \,\dvy\,,
\eq
and
\begin{eqnarray}
\frac{c\,\dEy}{B} = \dvx - \frac{1}{a}\left[\frac{\eta_P}{\eta_H}\left(1 - a + i\,\frac{k^2\,\eta_P}{\omega}\right)\right. \nonumber\\
+ \left. i\,\frac{k^2\,\eta_H}{\omega}\right]\,\dvy \,.
\end{eqnarray}
\section{Dispersion Relation}
We need four boundary conditions across $x = x_0$ in order to derive the dispersion relation. The first boundary condition, the continuity of the total pressure across the boundary $[\delta p_T] = 0$ gives 
\begin{eqnarray}
X_{in1}\,f_1\,\alpha_{in1}\,tanh\left(\alpha_{in1}\,x_0\right) + X_{in2}\, f_2\,\alpha_{in2}\,tanh\left(\alpha_{in2}\,x_0\right) \nonumber\\
= - \left( 
X_{ex1}\, s_1\,\alpha_{ex1} + X_{ex2}\, s_2\,\alpha_{ex2} 
\right)
\label{eq:bc1}
\end{eqnarray} 
where $X_{j} = Y_j + i\,Q_j\,G_{j}$ and 
\bq
Y_j = \beta \frac{\left(1 –- a_j\right)^2\,A_j}{a_j\,\Big[ 1 + \beta\,\left(1 -– a_j \right)\Big]}\,,  
\eq
and
\bq
Q_j = \frac{\eta_{P}}{a_j\,\eta_{H}}\left(1 –- a_j + i\,\frac{k^2\,\eta_{P}}{\omega}\right)  + i\,\left(\frac{\omega}{\omega_H}\right)\,. 
\eq
 Since Hall and Pedersen diffusivities in general are a function of ambient plasma parameters, they will have different values inside and outside the plasma slab. However, in order to keep the derivation tractable, while deriving dispersion relation we have assumed that the diffusivities are constant in the plasma.  

By writing the equations in the integral form (as a conservation law) for $\dvx$ we obtain $f_1 + f_2 = s_1 + s_2$. Third boundary condition is derived by writing the induction equation in the conservative form
\bq
\delt\B + \div \U = 0\,,
\label{eq:ind1}
\eq
where
\bq
\delta \U = \left(\delta\v + \delta \vB\right)\,\B - \B\,\left(\delta\v + \delta \vB\right)\,, 
\eq
which gives $[\delta \U] = 0$.  The final, fourth boundary condition is derived by demanding that the continuity of electric displacement across the surface \citep{Z96} $\left[\delta D_x \right] = 0$  where $\delta D_x \cong K_{xx} \dEx + K_{xy} \dEy$ with
\bq
K_{xx} \approx \frac{c^2}{\va^2}\,\,,K_{xy} \approx i\,\frac{c^2}{\va^2}\,\left(\frac{\omega}{\omega_H}\right)\,.
\eq
By imposing the above--mentioned boundary conditions and defining $S_1 = Q_{in} / Q_{ex}$, $S_2 = S_1 /\left(G_{ex1} -– G_{ex2}\right)$, we arrive at the following dispersion relation
\begin{eqnarray}
\left(\frac{\omega^2}{\oma_{in}^2} - 1\right)\,\,\left[ \frac{k^2}{q_{in}^2}\,T_i + i\,\frac{\eta_P}{\eta_H}\,\left(G_{in1}T_{in1} + G_{in2}T_{in2} \right)\right] 
\nonumber\\
+ S_2\,\left(\frac{\omega^2}{\oma_{ex}^2} - 1\right) 
 \left[\frac{k^2}{q_{ex}^2}\left(\alpha_{ex1}\,T_{ex2} - \alpha_{ex2}T_{ex1} \right) 
- i\,\frac{\eta_P}{\eta_H}\,N\right] 
\nonumber\\
- \left(\frac{\omega}{\omega_H}\right)\, \left(1 + \frac{\eta_P^2}{\eta_H^2}\right)\,\left[
G_{in1}T_{in1} + G_{in2}T_{in2}+ S_2\,N
\right]  = 0 \,,
\label{eq:dr1}
\end{eqnarray}
where 
\begin{eqnarray}
T_i = T_{in1} + T_{in2}\,,\\
T_{in1} = - S_3\,\alpha_{in1}\, 
\left(\begin{array}{c}\tanh\\
\coth
\end{array}\right)\left(\alpha_{in1}\,x_{in}\right)\,,
\nonumber\\
T_{in2} = \alpha_{in2} 
\left(\begin{array}{c}\tanh\\
\coth
\end{array}\right)\left(\alpha_{in2}\,x_{in}\right)\,,
\nonumber\\
T_{ex1} = G_{in2} - G_{ex1} / S_1 -– S_3\,\left(G_{in1} - G_{ex1} / S_1 \right)
 \,,
\nonumber\\
T_{ex2} = G_{in2} - G_{ex2} / S_1 -– S_3\,\left(G_{in1} - G_{ex2} / S_1 \right)\,,
\nonumber\\
N = \alpha_{ex1}\,G_{ex1}\,T_{ex2} - \alpha_{ex2}\,G_{ex2}\,T_{ex1}\,,
\end{eqnarray}
with
\bq
S_3 = \frac{1 -– C\, G_{in2}\,S_1}{1 -– C\,G_{in1}\,S_1}\,,
\quad
C =\left(\frac{\omega}{\omega_H}\right)^{-1}\, 
\frac{A_{2} - A_{1}}{\left(\rho_{in} / \rho_{ex} - 1 \right)}\,.
\eq
Here
\bq
A_1 = \frac{\rho_{in}\, Q_{ex}}{\rho_{ex}\, Q_{in}}\left(c_{1} - \left(\frac{\omega}{\omega_{Ain}}\right)^2\, c_{2}\right)\,,
\eq
and
\bq
A_2 = \left(c_{1} - \left(\frac{\omega}{\omega_{Aex}}\right)^2\, c_{2}\right)\,,
\eq
with
\begin{eqnarray}
c_1 = \left(\frac{\omega}{\omega_H}\right)^2
\left(1 + \frac{\eta_P^2}{\eta_H^2}\right)
+ i \,\frac{\eta_P}{\eta_H} \left(\frac{\omega}{\omega_H}\right)
\,,\nonumber\\
c_2 = 1 + i \,\frac{\eta_P}{\eta_H}\,\left(\frac{\omega}{\omega_H}\right)\,.
\end{eqnarray}

We shall note that when non—-ideal MHD effects are completely absent, i.e. setting $\eta_P = \eta_H = 0$ the dispersion relation Eq.~ (\ref{eq:dr1}) reduces to the following simple form
\bq
\left(\omega^2 - \omega_{Ain}^2\right) + R\sqrt{\frac{A_i}{A_e}}\left(\omega^2 - \omega_{Aex}^2\right)\tanh(\sqrt{A_i}\,K)= 0\,,
\label{eq:NH}
\eq    
which for $A_j = 1$ is Eq.~(11) of  \cite{E82} except now it pertains to the partially ionized medium. However, in the present case due to the compressibility effect, $A_j \neq 1$ and the above--mentioned dispersion relation is fairly complicated.  

There is no general prescription available to solve the dispersion relation, Eq.~(\ref{eq:dr1}). Therefore, we shall analyse it in various limiting cases. We first explore the role of Pedersen diffusion on the surface waves. To that end, we examine the dispersion Eq.~(\ref{eq:dr1}) in the long wavelength ($k\,x_0  \rightarrow 0$) limit. Defining $R = \rhox / \rhon$, $k x_0 \equiv K$, $H = L_H / x_0$, and normalized phase speed $V_{P} = \omega / \omega_{Ain} = 1 / \sqrt{a_{i}}$, we may write 
\begin{eqnarray}
A_{j} \approx - \frac{R_q\,V_P^2}{\beta}\,, F / F_1 \approx \frac{i\,\omega}{\eta_P}\,,\nonumber\\
Q_{in} \approx \frac{\eta_P}{\eta_H}\,V_{P}^2\,, Q_{ex} \approx R\,Q_{in}\,, C\approx i\,\frac{\eta_P}{\eta_H}\,R\,V_P^2 \,,
\nonumber\\
S_1 \approx  1/R\,,S_2 \approx –- 1 /G_{in2}\,,S_3 \approx 1 - \beta\,,
\nonumber\\
G_{in2} \approx \, G_{ex2} \approx -i\,\frac{\beta\,\eta_H}{ V_{P}^2\,\eta_P}\,,
\tanh(\alpha\,K) \approx \frac{i\,K\, V_{P}}{\sqrt{\beta}}\,,
\nonumber \\
T_{ex1} \approx  G_{in2}\,, T_{ex2} \approx \left(1 –-\beta\right)\,G_{in2}
\,,
\end{eqnarray}
where $R_{q} = 1\,,R$ for $q = in\,,ex$. 
With the above--mentioned approximations, the dispersion relation Eq.~(\ref{eq:dr1}) reduces to the following simple form
\begin{eqnarray}
V_P^3\Bigg\{-\left(1-\beta\right)K^2 + \frac{1}{2}\left(\frac{\eta_H}{\eta_P}\right)\,\frac{\left(K - 2\,i\right)\,\beta}{H\,V_P}\Bigg.
\nonumber\\
\Bigg.
- i\,\frac{\sqrt{R\,\beta}K\,\left(1-\beta\right)}{V_P}  
- \frac{\beta\,K}{2\,R\,V_P^2}\,\left(1 + i\,R\right)
\Bigg\}
\nonumber\\
+ \left(1-\beta\right)K^2\,V_P - \frac{1}{2}\left(\frac{\eta_H}{\eta_P}\right)\,\frac{K\,\beta}{H}
\nonumber\\
+ \frac{i}{R}\Bigg\{\sqrt{R\,\beta}K\,\left(1-\beta\right) + 
\frac{1}{2}\left(\frac{\eta_H}{\eta_P}\right)\frac{\beta}{H}
\Bigg\}= 0
\,.
\label{eq:drt}
\end{eqnarray}
Note that since in the first bracket, second and third terms are the dominant terms, we may balance it with the last term in the curly bracket above to yield
\bq
V_P \approx -– i\,\left(\frac{\eta_P}{\eta_H}\right)^{1/2}\,\sqrt{\frac{R\,H}{\sqrt{\beta}}\,\left(\beta-1\right)}\,,
\eq
which implies that the waves are damping at a rate $k\,\eta_P^{1/2}$. The damping rate is very similar to the well known viscous damping of surface waves in a compressible medium \citep{R86, R91, R00}. Clearly, the viscosity and Pedersen diffusion plays similar role at the interface. However, detailed physical mechanism in two cases are somewhat different. Whereas in the present case it is the ion magnetization (determined by a competition between the ion—-cyclotron frequency against the ion-—neutral collision) that is responsible for the damping of the wave, it could be anisotropic proton viscosity that may be responsible for the viscous damping of the waves \citep{R91}.       

We shall now analyse the dispersion relation for purely Hall case by setting $\eta_P = 0$. Note that
\bq
\left(\frac{\alpha^2}{A}\right) -– k^2 = - \left(\frac{F}{A\,F_1}\right)\,
\left(1 - \frac{k^2\,F_1}{a\,\beta\,F}\right)\,, 
\eq
and since
\bq
\frac{k^2\,F_1}{a\,\beta\,F} = k^2\,\left(\frac{\omega}{\omega_H}\right)^2\,\frac{a/(1-—a)}{\beta\,\left(1-a\right)} \ll 1\,,
\eq
in the long wavelength limit, we may write 
\bq
G_{j\,2} \simeq \frac{\eta_H}{A\,\omega}\,\frac{F}{F_1}\,\frac{1}{\left(1 –- a_j\right)}\,, 
\eq
for purely Hall case.  Defining $\veps = \omega / \omega_H$ and $\Delta = 1 + R $ we may write
\begin{eqnarray}
C = \frac{\Delta\, V_P^2}{\veps} - \veps\,,G_2 = \frac{V_P^2-1}{A\,\veps}\nonumber\\
S_3 = \frac{V_P^2}{A_i}\Bigg[1 - \frac{\Delta\,\left(V_P^2 - 1\right)}{\veps^2} - \frac{1}{\beta} \Bigg]\,.  
\end{eqnarray}
The dispersion relation Eq.~(\ref{eq:dr1}) becomes
\begin{eqnarray}
\left(V_P^2 -– 1\right) + \sqrt{\frac{A_i}{A_e}}\left(R\, V_P^2 -– 1\right)\tanh(\sqrt{A_i}\,K)  
\nonumber\\
 + \frac{\veps^2}{\Delta}\Bigg[ \frac{1}{\beta} - 1 + \sqrt{\frac{A_i}{A_e}}\,\left(\frac{R}{\beta} - \frac{1}{R}\right)
\tanh(\sqrt{A_i}\,K)\Bigg] 
= 0\,.
\label{eq:SHC}
\end{eqnarray}
In the incompressible ($\beta \rightarrow \infty$) limit Eq.~(\ref{eq:SHC}) becomes
\begin{eqnarray}
\left(\omega^2 - \omega_{Ain}^2\right) + R\,\left(\omega^2 - \omega_{Aex}^2\right) \tanh(K) 
\nonumber\\
 - \left(\frac{\omega}{\omega_H}\right)^2\,\omega_{Ain}^2\,\left[1 + \frac{1}{R}\tanh(K)\right] \left(1 + R \right)^{-1} = 0\,,
\label{eq:SIL}
\end{eqnarray}
which is Eq.~(59) \citep{P13} except for a typo in their last bracket.

\begin{figure}
\includegraphics[scale=0.30]{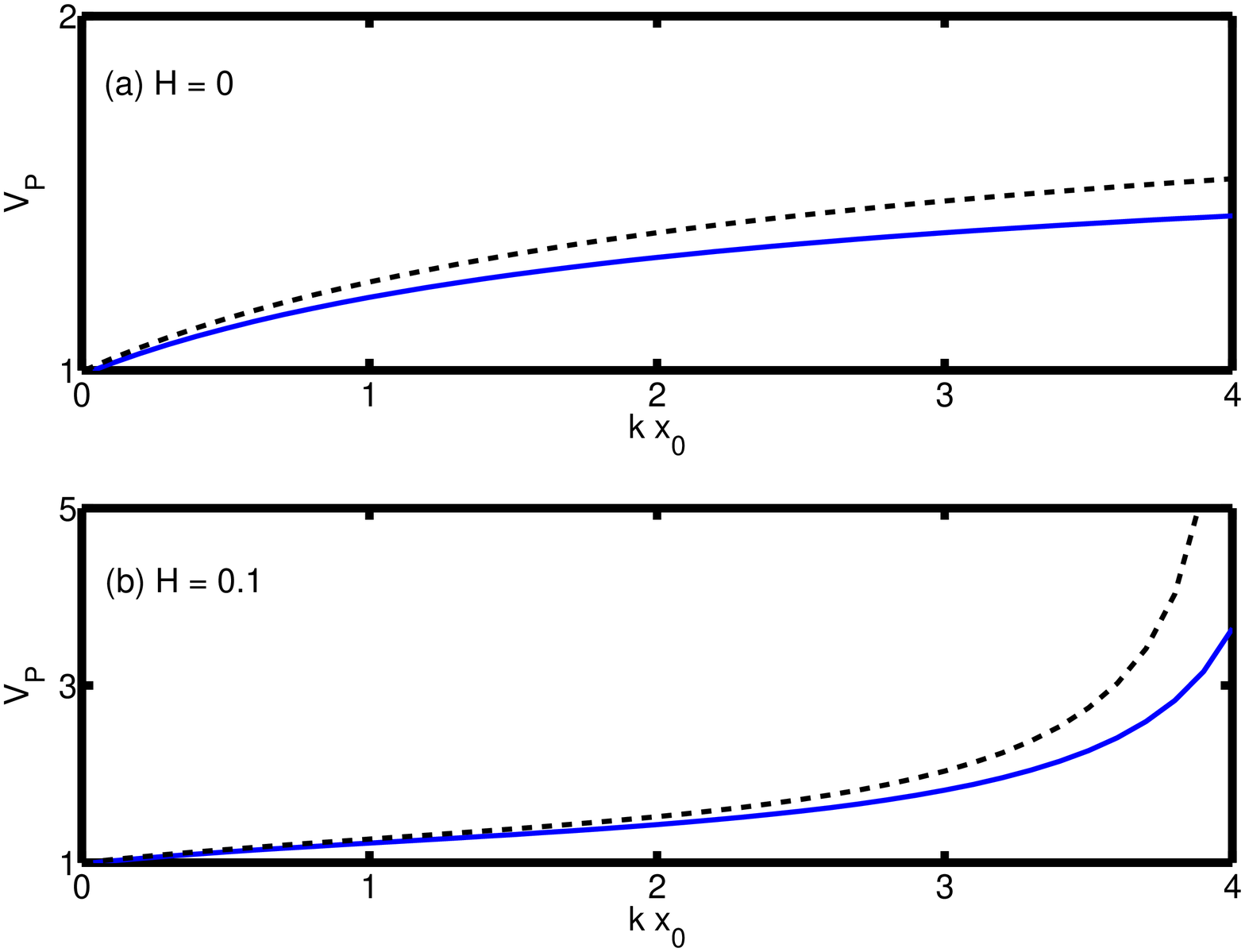}
\caption{\large{The normalized phase speed of kink wave $V_P$ against $k\,x_0$ for $\beta = 5$ (solid line) and $\beta = 10$ (dashed line) is shown in this figure with (frame b) and without (frame a) Hall diffusion.}}
\label{fig:fig2}
\end{figure}
We shall analyse the dispersion relation, Eq.~(\ref{eq:SHC}) by first noting that since $\sqrt{A_j}$ needs to be positive in order for kink wave to propagate in the medium, this implies  $V_P^2 / \beta < 1$. However, since the dispersion relation has been derived by assuming $\beta\,\left(1-—1/V_P^2\right)) \gg 1$, combining these two conditions gives $\beta \gg 1$. Thus approximating 
\bq
\sqrt{A} \approx 1 - \frac{V_P^2}{2\,\beta}\,,
\label{eq:aap}
\eq
and neglecting $V_P^4 / \beta$ terms, for a thin slab, the dispersion relation becomes
\bq
V_P^2 \approx \frac{2\,\beta\,\left(1+K\right)}{R\,\left(1+2\,\beta\,K\right) + 2\,\beta + K -– 2\,\beta\,H^2\,K^3 / R\,\left(1+R\right)}\,.
\label{eq:drH}
\eq    
In the absence of Hall ($H = 0$), the above--mentioned dispersion relation reduces to
\bq
V_P^2 \approx 1 + \left(1—-R\right)\,,
\eq
which is Eq.~(13) of \cite{E82} except here it describes the propagation of kink wave in partially ionized thin slab.  

We solve Eq.~(\ref{eq:drH}) by taking $R = 0.25$. The choice of $R$ is constrained by the equilibrium pressure balance across the slab. From Fig.~(\ref{fig:fig2}) we see that with increasing plasma $\beta$, the phase speed of the kink wave also increases. Further, in the absence of Hall [Fig.~(\ref{fig:fig2}(a)] the phase speed propagates at all $k\,x_0$, which is related to the generic nature of the ideal MHD like behaviour of the partially ionized medium at low frequency \citep{P13}. However, in the presence of Hall [Fig.~(\ref{fig:fig2}(b)], which incidentally introduces a scale $L_H$ in the system, the waves may propagate in the medium only if
\bq
\frac{H^2\,K^2}{ R\,\left(1+R\right)} < R\,\left(1 + \frac{1}{2\,\beta\,K}\right) + \frac{1}{K} + \frac{1}{2\,\beta}\,.
\label{eq:tt1}
\eq      
 It is clear that the cut—-off introduced by Hall scale is only weakly dependent on the plasma $\beta$ since Eq.~(\ref{eq:drH}) has been derived in $\beta \gg 1$ limit. 

In thick slab case , when $\sqrt{A_{in}}K \gg 1$, after approximating $\sqrt{A}$ as in Eq.~(\ref{eq:aap}), the dispersion relation Eq.~(\ref{eq:SHC}) gives
\bq
V_P^2 \approx \frac{1}{1 + R - \frac{H^2\,K^2}{R}}\,,
\label{eq:tte}
\eq
which to the leading order, has no dependence on compressibility. Further, we note that this is Eq.~(63) of \cite{P13} except for the absence of $1/R$ in their Hall term. This is due to the fact that the Hall term in their dispersion relation Eq.~(59) contains $1/(1 + 1/R)$ rather than $1/(1 +R)$. In the absence of Hall, above equation reduces to Eq.~(14) of \cite{E82}.  

\begin{figure}
\includegraphics[scale=0.30]{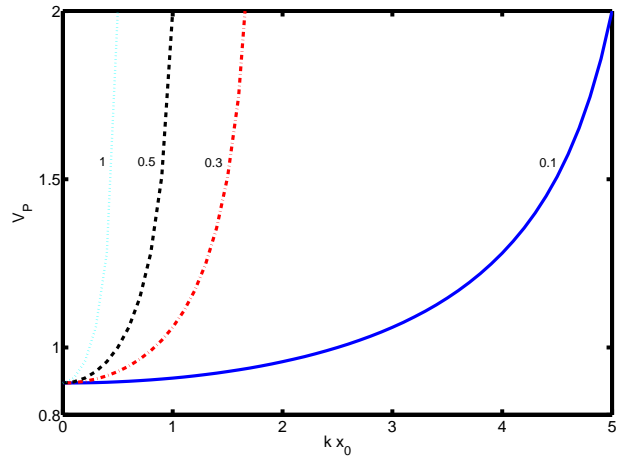}
\caption{\large{The normalized phase speed of kink wave $V_P$ against $k\,x_0$ for $L_H = \left(0.1\,,0.3\,,0.5\,,1\right)\,x_0$  is shown in this figure.}}
\label{fig:fig3}
\end{figure}

In Fig.~(\ref{fig:fig3}), we plot Eq.~(\ref{eq:tte}) for $R = 0.25$. It is clear that like thin slab, Hall diffusion introduces a cut—-off in the thick slab as well. Only waves with wavelength larger than
\bq
\lambda > \frac{2\,\pi\,L_H}{\sqrt{R\,\left(1+R\right)}}\,,
\label{eq:tt2}
\eq
can propagate in a thick slab. Clearly, Hall scale, which is an artefact of neutral-—ion collision in a weakly ionized medium \citep{P06, P08a} introduces such a cut—-off. 

\section{Discussion and Summary}
How compressibility does affect the wave propagation in the solar photosphere—-chromosphere region? 
\begin{figure}
\includegraphics[scale=0.30]{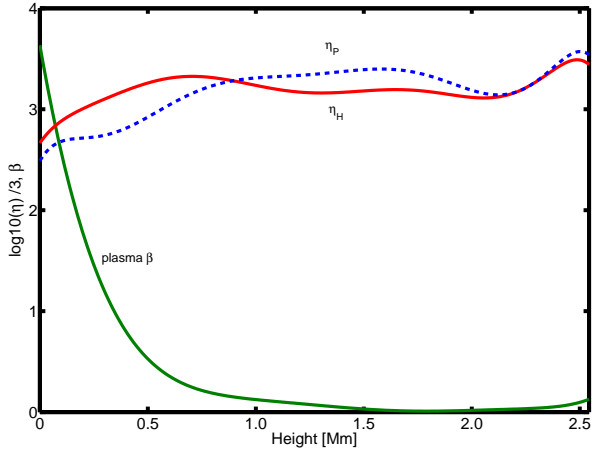}
\caption{\large{The height variation of the Pedersen and Hall diffusivities along with the plasma $\beta$ is shown in this figure.}}  
\label{fig:fig4}
\end{figure}
To answer this, we plot in Fig.~(\ref{fig:fig4}) plasma $\beta$ along with the magnetic diffusivities by assuming a flux  profile 
\bq
B = B_0\,\left(\frac{n_n}{n_i}\right)^{(0.3)}\,\mbox{G}\,,
\eq
where $B_0 = 1200\,\mbox{G}$. The temperature and density is taken from Model C (table 12) of \cite{VAL81}. The Pedersen and Hall diffusion profiles have been smoothed by fitting a polynomial to actual model data. It is clear from Fig.~(\ref{fig:fig3}) that in the large part of photosphere, plasma $\beta \gg 1$ whereas in the upper photosphere ($\sim 0.5\,\mbox{Mm}$) and beyond, $\beta \lesssim 1$. Clearly, in the middle and upper chromosphere, incompressibility is a poor assumption. However, this is only valid if the field is $1\,\mbox{kG}$ or, more at the footpoint. For a weaker field, e.g. when $B_0 = 120\,\mbox{G}$ at the footpoint, since the value of $\beta$ shown in the figure will be multiplied by a factor $100$, the plasma can be treated as incompressible throughout the entire photosphere--chromosphere. Thus we infer from Fig.~(\ref{fig:fig2}) that the kink waves in thin slab will have higher phase speed in the weak field regions. Therefore, intergranular boundaries of the photosphere (known for strong vertical fields; \citep{SL64}) will launch less energetic waves than the internetwork regions which often are sites of weaker field. It is pertinent to note here that isolated intense field have also been detected in the internetwork regions \citep{D09, SZ10}. Thus it is quite possible that kink wave of varying energy is excited in both network and internetwork regions.  

Note that the Hall diffusion introduces a cut—-off, which is similar in both thin and thick slabs. Since the Hall scale is typically a few km in the photosphere and lower and middle chromosphere \citep{P13}, this suggests that waves with wavelength larger than $30—-40$ km will propagate in  plasma slab.  The Hall cut—-off is not very sensitive to the plasma beta.  However, Fig.~(\ref{fig:fig4}) suggests that low--frequency sausage and kink wave may not survive in the upper chromosphere altogether. The Pedersen diffusion (which dominates Hall in the middle and upper chromosphere) will damp these waves whose severity will depend on the wave amplitude and ambient plasma parameters.     

We shall note that the present model suffers from similar limitations as our previous incompressible model \citep{P13}. Namely, piecewise constant density and pressure profile is an idealization and do not capture the resonant behaviour of the \alf surface waves \citep{G13}. Even with this limitation, the model becomes quite complex to handle analytically. Owing to this complexity we have analysed only very simple $\beta \gg 1$ case. As noted above, plasma beta may become small 
in the network regions and only recourse to investigate the effect of compressibility  on wave propagation in these regions may be numerical.

To summarize, wave propagation in partially ionized plasma slab only weakly depends on the plasma compressibility. For example waves launched at intergranular boundaries may have somewhat larger phase speed than waves emanating from internetwork region. However, more than the compressibility of the medium, it is the fractional ionization and associated magnetic diffusivities that determine the characteristics of normal modes in slabs.

Following is the summary of this work.

1. The partially ionized solar photosphere-chromosphere plasma is incompressible in the internetwork region whereas, at intergranular boundaries plasma compressibility varies with the scale height.  

2. The surface waves in the compressible plasma medium only weakly depends on the plasma $\beta$. The phase velocity of sausage and kink waves marginally (by a few percent) increases due to the compressibility of a partially ionized medium in both {\it ideal} as well as Hall diffusion dominated regimes. 

3. Unlike {\it ideal} case, in the Hall--dominated regime only waves below certain cut—-off frequency can propagate in the medium. This cut-—off for a thin slab has a weak dependence on the plasma compressibility whereas for thick slab no such dependence exists. More importantly, since the cut-—off is introduced by the Hall diffusion, the fractional ionization of the medium is more important than plasma compressibility in determining such a cut—off.    
  
4. The ambipolar diffusion always damps long wavelength fluctuations and is independent of plasma compressibility. 

\section*{Acknowledgements}
 BP wishes to acknowledge the financial support of the Australian Research Council through grant DP 130104873. This research has made use of NASA$\textquoteright$s Astrophysics Data System.


\begin{thebibliography}{}
\bibitem[\protect\citeauthoryear{Aschwanden}{2009}]{A09}
Aschwanden, M., 2009, Phyics of the Solar Corona, Sringer—Praxis, Chichester
\bibitem[\protect\citeauthoryear{Braginskii}{1965}]{B65}
Braginskii, S. I., 1965, in Leontovich M. A., Review of Plasma Physics, Vol. 2, Consultants Bureau, New York, p. 205
\bibitem[\protect\citeauthoryear{Cally}{1985}]{C85}
Cally, P. S., 1985, Aust. J. Phys., 38, 825
\bibitem[\protect\citeauthoryear{Cally}{1986}]{C86}
Cally, P. S., 1986, Sol. Phys., 103, 277
\bibitem[\protect\citeauthoryear{Chen \& Hasegawa}{1974}]{C74}
Chen, L. \& Hasegawa A., 1974, Phys. Fluids, 17, 1399
\bibitem[\protect\citeauthoryear{Cowling}{1957}]{C57}
Cowling, T. G., 1957, Magnetohydrodynamics, Adam Hilger, Bristol
\bibitem[\protect\citeauthoryear{de Wijn et al.}{2009}]{D09}
de Wijn A. G., Stenflo, J. O., Solanki, S. K. \& Tsuneta S., 2009, Space Sci. Rev., 144, 275
\bibitem[\protect\citeauthoryear{Edwin \& Roberts}{1982}]{E82}
Edwin P. M. \& Roberts B., 1982, Sol. Phys., 76, 239
\bibitem[\protect\citeauthoryear{Goodman}{1998}]{G98}
Goodman, M. L., 1998, ApJ, 503, 938
\bibitem[\protect\citeauthoryear{Goossens}{1994}]{G94}
Goossens, M., 1994, Space Sci. Rev., 68, 51
\bibitem[\protect\citeauthoryear{Goossens et al.}{2009}]{G09}
Goossens, M., Terradas, J., Andries, J., Arregui, I. \& Ballester, J. L., 2009, \AnA, 503, 213
\bibitem[\protect\citeauthoryear{Goossens et al.}{2011}]{G11}
Goossens, M., Erdelyi, R. \& Ruderman, M. S., 2011, Space Sci. Rev., 158, 289
\bibitem[\protect\citeauthoryear{Goossens et al.}{2012}]{G12}
Goossens, M., Andries, J., Soler, R., Van Doorsselaere, T., Arregui, I. \& Terradas, J., 2012, \ApJ, 753, 111
\bibitem[\protect\citeauthoryear{Goossens et al.}{2013}]{G13}
Goossens, M., Van Doorsselaere, T., Soler \& Verth, G., 2013, \ApJ, 768, 191
\bibitem[\protect\citeauthoryear{Ionson}{1978}]{I78}
Ionson, J. A., 1978, ApJ, 226, 650 
\bibitem[\protect\citeauthoryear{Kapparaff \& Tataronis}{1977}]{K77}
Kapparaff, J. M. \& Tataronis, 1977, J. A., J. Plasma Phys., 18, 209
\bibitem[\protect\citeauthoryear{Kazeminezhad \& Goodman}{2006}]{KG06}
Kazeminezhad, F. \& Goodman, M. L., 2006, \ApJS, 166, 613
\bibitem[\protect\citeauthoryear{Khodachenko \& Zaitsev}{2002}]{K02}
Khodachenko, M. L.\& Zaitsev, V. V., 2002, Ap\&SS, 279, 389.
\bibitem[\protect\citeauthoryear{Khodachenko et al.}{2006}]{K06}
Khodachenko, M. L., Rucker, H. O., Oliver, R., Arber, T. D.\& Hanslmeier, 2006, Adv. Space Res., 37, 447.
\bibitem[\protect\citeauthoryear{Khomenko \& Collados}{2012}]{K12}
Khomenko E. \& Collados, M., 2012, \ApJ, 747, 87.
\bibitem[\protect\citeauthoryear{Khomenko et al.}{2014}]{K14}
Khomenko E., Collados, M., Diaz, A. \& Vitas N., 2014, Phys. Plasmas,21, 092901 
\bibitem[\protect\citeauthoryear{Mett \& Taylor}{1992}]{M92}
Mett, R. R. \& Taylor, J. B., 1992, Phys. Fluids B, 4, 73 
\bibitem[\protect\citeauthoryear{Pandey}{2013}]{P13}
Pandey B. P., 2013, MNRAS, 436, 1659 
\bibitem[\protect\citeauthoryear{Pandey \& Vladimirov}{2007}]{PV07}
Pandey B. P. \& Vladimirov, S. V., 2007, ApJ, 664, 942
\bibitem[\protect\citeauthoryear{Pandey \& Wardle}{2006}]{P06}
Pandey B. P. \& Wardle, M., 2006, preprint (astroph/0608008v2)
\bibitem[\protect\citeauthoryear{Pandey \& Wardle}{2008}]{P08a}
Pandey B. P. \& Wardle, M., 2008, MNRAS, 385, 2269 
\bibitem[\protect\citeauthoryear{Pandey \& Wardle}{2012}]{PW12}
Pandey B. P. \& Wardle M., 2012, MNRAS, 426, 1436 
\bibitem[\protect\citeauthoryear{Pandey \& Wardle}{2013}]{PW13}
Pandey B. P. \& Wardle, M., 2013, MNRAS, 431, 570 
\bibitem[\protect\citeauthoryear{Pandey et al.}{1995}]{P95}
Pandey B. P., Avinash, K., Kaw, P. K. \& Sen, A., 1995, Phys. Plasmas, 2, 629 
\bibitem[\protect\citeauthoryear{Pandey et al.}{2008}]{P08b}
Pandey B. P., Vranjes, J. \& Krishan, V., 2008, MNRAS, 386, 1635 
\bibitem[\protect\citeauthoryear{Parhi et al.}{1997b}]{P97a}
Parhi, S., Pandey, B. P., Lakhina G. S, Goossens, M. \& de Bruyne P., 1997, AdSpR, 19, 1891
\bibitem[\protect\citeauthoryear{Parhi et al.}{1997a}]{P97b}
Parhi, S., Pandey, B. P., Goossens, M., Lakhina G. S., de Bruyne P.,1997, Ap\&SS, 250, 147


\bibitem[\protect\citeauthoryear{Parhi et al.}{1998}]{P98}
Parhi, S., Pandey, B. P., Goossens, M. \& Lakhina, G. S., 1998, in Kurtz D. W., Christensen-—Dalsgaard J., eds, Proc. IAU Symp. 185, New Eyes to See Inside the Sun and the Stars: Pushing the Limits of Helio and Astero-—Seism. Kluwer, Dordrecht, P.457  
\bibitem[\protect\citeauthoryear{Parker}{1972}]{P72}
Parker, E. N., 1972, Sol. Phys., 37, 127
\bibitem[\protect\citeauthoryear{Roberts \& Webb}{1979}]{R79}
Roberts, B. \& Webb, A. R. 1979, Sol. Phys., 64, 77-92
\bibitem[\protect\citeauthoryear{Ruderman}{1986}]{R86}
Ruderman, M. S., 1986, Fluid Dyn., 21, 925
\bibitem[\protect\citeauthoryear{Ruderman}{1991}]{R91}
Ruderman, M. S., 1991, Sol. Phys., 131, 11
\bibitem[\protect\citeauthoryear{Ruderman et al.}{2000}]{R00}
Ruderman, M. S., Oliver, R., Erd\'elyi, R., Ballester, J. L. \& Goossens, M., 2000, \AnA, 354, 261
\bibitem[\protect\citeauthoryear{Sanchez et al.}{2010}]{SZ10}
Sanchez J. S., Bonet, J. A., Viticchi\'es, B. \& Del Moro, D., 2010, \ApJ, 715, L26
\bibitem[\protect\citeauthoryear{Simon \& Leighton}{1964}]{SL64}
Simon G. W. \&  Leighton R. B., 1964, \ApJ, 140, 1120
\bibitem[\protect\citeauthoryear{Soler at al.}{2009}]{S09}
Soler, R., Oliver, R., \& Ballester, J. L., 2009, \ApJ, 699, 1553 
\bibitem[\protect\citeauthoryear{Soler at al.}{2010}]{S10}
Soler, R., Oliver, R., \& Ballester, J. L., 2010, \AnA, 512, A28 
\bibitem[\protect\citeauthoryear{Soler at al.}{2013}]{S13}
Soler, R., Oliver, R., \& Ballester, J. L., 2013, \ApJ, 767, 171 
\bibitem[\protect\citeauthoryear{Tataronis \& Grossman}{1973}]{T73}
Tataronis, J. A. \& Grossman, W., 1973, Nucl. Fusion, 16, 667
\bibitem[\protect\citeauthoryear{Vernazza et al.}{1981}]{VAL81}
Vernazza J. E., Avrett E. H. \& Loser, R., 1981, \ApJS, 45, 635
\bibitem[\protect\citeauthoryear{Uberoi \& Datta}{1998}]{U98}
Uberoi C. \& Datta A., 1998, Phys. Plasmas, 5, 4149
\bibitem[\protect\citeauthoryear{Vranjes et al.}{2008}]{V08}
Vranjes, J., Poedts, S., Pandey, B. P. \& De Pontieu, B., 2008, \AnA, 478, 553 
\bibitem[\protect\citeauthoryear{Wardle}{1999}]{W99}
Wardle M., 1999, MNRAS, 307, 849
\bibitem[\protect\citeauthoryear{Wentzel}{1979}]{W79}
Wentzel D. G. 19799, ApJ, 233, 756 
\bibitem[\protect\citeauthoryear{Zaqarashvili et al.}{2012}]{Z12}
Zaqarashvili, T. V., Carbonell, M., Ballester, J. L. \& Khodachenko, M. L., 2012, \AnA, 3, 4346
\bibitem[\protect\citeauthoryear{Zhelyazkov et al.}{1996}]{Z96}
Zhelyazkov, I., Debosscher, A. \& Goossens, M., 1996, Phys. Plasmas, 3, 4346
\bibitem[\protect\citeauthoryear{Zhelyazkov}{2009}]{Z09}
Zhelyazkov, I., 2009,  Eur. Phys. J. D, 55, 127
\end{thebibliography}
\end{document}